\documentclass[11pt]{article}
\usepackage{amssymb,amsbsy,amsmath,amsfonts,amssymb,amscd}
\newtheorem{e-proposition}[theorem]{Proposition}

\newtheorem{e-definition}[theorem]{D\'efinition\rm}

\newcommand{\Rm}{{\mathbb R}}
\newcommand{\eps}{\varepsilon}

\newcommand{\dint}{\displaystyle\int}
\newcommand{\bx}{{\bf x}}
\newcommand{\vx}{{\bf x}}
\newcommand{\vz}{{\bf z}}
\newcommand{\vy}{{\bf y}}
\newcommand{\vb}{{\bf b}}
\newcommand{\by}{{\bf y}}
\newcommand{\bz}{{\bf z}}
\newcommand{\bk}{{\bf k}}
\newcommand{\bp}{{\bf k}}
\newcommand{\bq}{{\bf q}}
\newcommand{\bu}{{\bf u}}
\newcommand{\bv}{{\bf v}}

\newcommand{\Bxi}{\boldsymbol{\xi}}

\def\bS{{\bf S}}

\newcommand{\pdr}[2]{\dfrac{\partial #1}{\partial #2}}

\begin{document}
\title{%
Time Reversal for Classical Waves in Random Media
}
\author{Guillaume Bal and Leonid Ryzhik
\\
  Department of Mathematics, University of Chicago,\\
  Chicago IL 60637\\
gbal@math.uchicago.edu, ryzhik@math.uchicago.edu}

\vspace{2cm}

\maketitle

\thispagestyle{empty}
\noindent \begin{tabular}{@{} p{2cm} @{} p{12cm} @{} }
{\bf Abstract.} & {\small
We propose a mathematical theory
for the refocusing properties observed in time-reversal 
experiments, where classical waves propagate through a medium,
are recorded in time, then time-reversed and sent back into the medium.
The salient feature of such experiments is that
the refocusing quality of the time-reversed reemitted signals is greatly
enhanced when the underlying medium is heterogeneous. Based on the
Wigner transform formalism, we show that random media indeed greatly
improve refocusing. We analyze two different types of random media,
where in the limit of high frequencies, the Wigner transform satisfies
a random Liouville equation or a linear transport equation.
}
\\
\vspace{.1cm}
\\
 & {\bf \large Renversement du temps pour la propagation d'ondes
  classiques  en milieu al\'eatoire} 
\\
\vspace{0cm}
\\
{\bf R\'esum\'e.} & {
    Dans des exp\'eriences de renversement temporel, o\`u un signal
    acoustique qui se propage dans un milieu sous-jacent, est
    enregistr\'e en temps, puis subit un renversement temporel et est
    r\'e\'emis dans le milieu, il a \'et\'e observ\'e que le signal
    refocalise d'autant mieux \`a l'emplacement de la source initiale
    que le milieu est h\'et\'erog\`ene. Nous donnons une explication
    math\'ematique de ce ph\'enom\`ene fond\'ee sur le formalisme de
    la transform\'ee de Wigner. Nous obtenons deux exemples de
    refocalisation, en fonction du type de milieu al\'eatoire
    consid\'er\'e, selon que la transform\'ee de Wigner, \`a la limite
    des hautes fr\'equences, satisfait soit une \'equation de
    Liouville avec coefficients al\'eatoires, ou une \'equation de
    transport.  
}

\end{tabular}

\vspace{.0cm}

\section*{Version fran\c caise abr\'eg\'ee}

Dans cette note, nous nous int\'eressons \`a la propagation
d'ondes classiques en milieu al\'eatoire et \`a leurs propri\'et\'es
de refocalisation apr\`es renversement temporel. Dans les
exp\'eriences conduites par M. Fink \cite{Fink-PT} et ses
collaborateurs, une source localis\'ee \'emet des ondes acoustiques
\`a travers un milieu h\'et\'erog\`ene.  Le signal est enregistr\'e en
temps par des capteurs localis\'es en espace et positionn\'es loin de
la source, puis renvers\'e en temps et enfin retransmis dans le milieu
(ce qui est arriv\'e en dernier repart en premier).

Dans ces exp\'eriences, il a \'et\'e observ\'e le ph\'enom\`ene
suivant, qui \`a premi\`ere vue pourrait para{\^{\i}}tre
contre-intuitif~: le signal r\'e\'emis refocalise \`a l'emplacement de
la source initiale d'autant mieux que le milieu est h\'et\'erog\`ene,
alors que la qualit\'e de la refocalisation est tr\`es mauvaise en
milieu homog\`ene. L'explication r\'eside dans la multiplicit\'e des
``chemins'' menant de la source aux capteurs. Plus le milieu est
al\'eatoire, plus il donne d'informations sur la source aux capteurs,
pourvu que les enregistrements soient suffisamment longs.

Plusieurs travaux sur ce ph\'enom\`ene existent dans la litt\'erature
physique \cite{DJ90,Fink-PT,Fink-Prada-01}. Les r\'esultats
math\'ematiques existants portent sur le cas mono-dimensionnel
\cite{Fouque-Clouet} et le r\'egime de faisceaux \'etroits
\cite{BPZ-JASA01}. Dans cette note, nous proposons une th\'eorie qui
explique le ph\'enom\`ene de refocalisation pour des ondes classiques
\`a hautes fr\'equences en milieu multi-dimensionnel.  Notre approche,
s'inspirant de \cite{BPZ-JASA01}, se fonde sur la transform\'ee de
Wigner, d\'efinie en (\ref{eq:wignertfsm}), qui permet l'analyse des
corr\'elations de noyaux de Green en des points voisins. Nous pouvons
\'ecrire le signal retransmis apr\`es renversement du temps en
fonction de cette transform\'ee de Wigner (\ref{eq:psiBeps2sm}).

Selon les caract\'eristiques du milieu al\'eatoire, la transform\'ee
de Wigner satisfait diff\'erentes \'equations limites quand la
fr\'equence caract\'eristique des ondes tend vers l'infini.  Nous
analysons les deux cas o\`u nous obtenons \`a la limite soit une
\'equation de Liouville (limite semiclassique), soit une \'equation de
transport de type Boltzmann (limite de couplage faible). Dans le
premier cas, nous sommes capables d'obtenir une d\'erivation
rigoureuse math\'ematiquement. La formule de refocalisation est
donn\'ee en (\ref{eq:reconst})-(\ref{eq:FLt}). Dans le second cas, la
d\'erivation de la formule de reconstruction
(\ref{eq:reconst})-(\ref{eq:Ft}) est formelle en certains endroits.
La th\'eorie permettant d'obtenir l'\'equation de transport limite
n'est compl\`ete que pour l'\'equation (scalaire) de Schr\"odinger
\cite{Erdos-Yau2,Spohn}. De plus elle n\'ecessite de moyenner sur les
r\'ealisations du milieu al\'eatoire, bien que les exp\'eriences
physiques montrent que ce soit superflu.

\par\medskip\centerline{\rule{2cm}{0.2mm}}\medskip
\setcounter{section}{0}

\section*{\large 1. Refocusing in time-reversal}

Time-reversal experiments \cite{Fink-PT} can be briefly
described as follows. A point source signal is sent through an
inhomogeneous medium and is recorded by a spatially localized array of
receivers-transducers. The signal is subsequently reversed in time and
reemitted back into the medium; that is, the part of the signal that
was recorded last is sent back first.  The refocusing of the reemitted
signal is then observed on the spot of the original source.

Refocusing in a homogeneous medium is poor when only a few
receivers are used to record the signal.  
The most striking and
somewhat counterintuitive observation is that inhomogeneities in the
medium {\em enhance} the refocusing effects.  This is because waves
recorded at the time-reversal array have sampled a larger part of the
medium than in the homogeneous case and carry more information about
the {\em source} location. Furthermore, the refocused signal in a
random medium is self-averaging, that is, independent of the
realization of the random medium.

Many time-reversal experiments have been conducted in the recent past
and studied theoretically \cite{DJ90,Fink-PT,Fink-Prada-01}.
Mathematical studies have concentrated so far on the one space
dimension case \cite{Fouque-Clouet} and on the narrow beam
approximation \cite{BPZ-JASA01,PRS}.  In this note we present an
analysis of the enhanced refocusing property in genuinely three
dimensional heterogeneous media in the regimes of random geometrical
acoustics and radiative transport. Our main tool is the asymptotic
analysis of the equation for the Wigner transform \cite{GMMP,LP}.  It
turns out that in the high frequency limit the refocused signal may be
written as a convolution with a kernel $F$ expressible in terms of the
limit Wigner distribution. More precisely, this kernel is expressed
(\ref{eq:kernelF}) in terms of the Fourier transform in $\bk$ of the
energy densities $a(t,\vx,\bk)$ in the phase space
(\ref{wig-decomp}). The latter are smooth in $\bk$ under evolution in
random media but not in a homogeneous domain. Therefore the spatial
localization of the kernel $F$ is improved in a random medium compared
to the homogeneous case. In the {\em geometric acoustics} limit, the
Wigner transform asymptotically satisfies a Liouville or a
Fokker-Plank equation similar to the one used to describe the
refocusing of narrow beam signals \cite{BPZ-JASA01,PRS}.  In the {\em
radiative transfer} regime, the Wigner distribution satisfies a linear
Boltzmann equation.  The rigorous passage to this limit is not
complete for classical waves
\cite{Erdos-Yau2,Ho-Landau-Wilkins,RPK-WM,Spohn} and our results in
this part remain formal.

\section*{\large 2. Reemitted signal and the Wigner transform}

The propagation of many classical waves is described by 
systems of $m$ first-order differential equations of the form
(repeated indices are summed over throughout the paper)
\begin{equation}
  \label{eq:syst} A(\bx) \pdr{\bu(t,\bx)}{t} + D^j
  \pdr{\bu(t,\bx)}{x_j}=0,¸~~\bx\in{\mathbb R}^d, d=2\hbox{ or } 3
\end{equation}
with initial conditions $\bu(0,\bx)=\bu_0(\bx)$.  In this paper, we
address time-reversal for acoustic waves with $d=3$ to fix notation.
The corresponding theory for other types of waves will be presented
elsewhere \cite{BPRS}. In this context, $\bu$ is the 4-vector composed
of the three components of the velocity field $\bv$ and the scalar
pressure field $p$. The matrix $A(\bx)={\rm
  Diag}(\rho(\bx),\rho(\bx),\rho(\bx),\kappa(\bx))$, where $\rho(\bx)$
is the density of the underlying medium and $\kappa(\bx)$ its
compressibility.  The $4\times 4$ matrices $D^j$ have entries
$D_{ml}^j=\delta_{m4}\delta_{lj}+\delta_{l4}\delta_{mj}$.  The
solution of (\ref{eq:syst}) is
\begin{math}
  \label{eq:decomp}
  \bu(t,\bx)=\int_{\Rm^d} G(t,\bx;\by) \bu_0(\by) d\by.
\end{math}
Here $G(t,\bx;\by)$ is the matrix-valued Green's propagator from
$\by$ to $\bx$. It  solves (\ref{eq:syst})
with the initial condition $G(0,\bx;\by)=I\delta(\bx-\by)$ where $I$ is
the $4\times 4$ identity matrix. Our time reversal setting is as
follows. At time $t$ the wave field is truncated in a region $\Omega$,
where the transducers-receivers are located.
Then the direction of the acoustic velocity field is reversed and the
signal is re-emitted into the medium. The back-propagated signal
emanating from the source term $\bu_0$ at time $0$ can be written
using the Green's propagator as
\begin{equation}
  \label{eq:psiB}
  \bu^B(\bx)= \dint_{\Rm^{3d}} \Gamma G(t,\bx;\by) \Gamma G(t,\by';\bz)
               \chi_\Omega(\by) \chi_\Omega(\by') f(\by-\by')
               \bu_0(\bz) d\by d\by' d\bz.
\end{equation}
Here $\Gamma$ is a matrix that models the time reversal process. 
It is given by 
$\Gamma=\rm{Diag}(-1,-1,-1,1)$, so that the velocity field $\bv$ is
replaced by $-\bv$ and the pressure field $p$ remains unchanged. 
The function
$\chi_\Omega(\by)$ may equal $1$ on $\Omega$ and $0$ outside, or
could be a more general function modeling amplification of the
time-reversed signal by various receivers.  We also allow for some
blurring of the recorded signal before it is transmitted back. The
convolution with a filter $f$
describes this blurring.  Introducing the adjoint Green's matrix
$G_*$, solution of
\begin{equation}
  \label{eq:greenadj}
  \begin{array}{l}
   \pdr{G_*(t,\bx;\by)}{t} + \pdr{}{x_j}(G_*(t,\bx;\by)) D^j A^{-1}(\bx)=0, 
  \end{array}
\end{equation}
with initial condition $G_*(0,\bx;\by)=\delta(\bx-\by) A^{-1}(\bx)$,
we observe that $\Gamma G_*(t,\bx;\by) A(\bx) \Gamma = G(t,\by;\bx)$.

We now rescale our problem. Refocusing is expected to be important in
the close vicinity of the source location. We take the initial source
as a localized function $\bu_0(\vx)={\bf
S}\left(\frac{\vx-\vx_0}{\eps}\right)$ of finite amplitude and rescale
the filter accordingly:
$\frac{1}{\eps^d}f(\frac{\vy-\vy'}{\eps})$. Here $\eps\ll 1$ is a
small parameter that measures the ratio of the width of the initial
pulse to the propagation distance $L$ between the source location and
the array of receivers-transducers.  An observation point $\vx$ is
close to $\vx_0$ and we write it as $\vx=\vx_0+\eps\Bxi$. After a
change of variables, equation (\ref{eq:psiB}) becomes
\begin{displaymath}
    \bu^{B}_{\eps}(\Bxi; \bx_0)\!=\!\!\dint\limits_{\Rm^{3d}} 
\Gamma   G(t,\bx_0+\eps\Bxi;\by)
    G_{*}(t,\bx_0+\eps\bz;\by') A(\bx_0+\eps\bz)
       \Gamma {\bf S}(\bz) \chi_\Omega(\by) \chi_\Omega(\by')
       f(\dfrac{\by-\by'}{\eps})d\by d\by'd\bz.
\end{displaymath}
The {\em Wigner transform} is a
natural tool in the analysis of the correlation function of wave
fields at neighboring points \cite{GMMP,LP,RPK-WM}.
In our context, we define it as
\begin{equation}
  \label{eq:wignertfsm}
  \!\!\! W_\eps(t,\vx,\bk)\!= \!\!
\dint_{\Rm^{2d}} \!\left[
\int_{\Rm^d} \!\!\!e^{i\bp\cdot\vz}G(t,\vx-\frac{\eps\vz}{2};\by)
G_{*}(t,\vx+\frac{\eps\vz}{2};\vy')\frac{d\vz}{(2\pi)^d}\right] \!
\chi_\Omega(\by)\chi_\Omega(\by') f(\dfrac{\by-\by'}{\eps}) d\by d\by'.
\end{equation}
This allows us to recast the expression for the retransmitted signal
as follows
\begin{equation}
  \label{eq:psiBeps2sm}
  \bu^{B}_{\eps}(\Bxi; \bx_0)= \dint_{\Rm^{2d}}
       \Gamma W_\eps(t,\bx_0+\eps\dfrac{\Bxi+\bz}{2},\bk)
      e^{-i\bk\cdot(\bz-\Bxi)}A(\bx_0+\eps\bz)\Gamma \bS(\bz) d\bz d\bk.
\end{equation}
Let us define the space ${\mathcal A}(\Rm^{2d})$ as
the subset of ${\mathcal S}'(\Rm^{2d})$ of matrix-valued 
distributions $\eta(\bx,\bk)$ such
that $\int_{\Rm^d} \sup_\bx \|\hat \eta(\bx,\by)\|d\by$ is bounded. Here
$\hat \eta(\bx,\by)$ is the inverse Fourier transform of $\eta(\bx,\bk)$ in
the second variable only.  We denote by ${\mathcal A}'$ the dual space to
${\mathcal A}$. Then we have
\\
\vspace{-.2cm} \\
\noindent {\bf Lemma 1.} {\em The Wigner transform $W_\eps(t,\bx,\bk)$
  is bounded in ${\mathcal C}^0(0,T;{\mathcal A}'(\Rm^{2d}))$ independent of
  $\eps$ provided that $\hat f(\bk)\in L^1({\mathbb R}^d)$. As a
  consequence, it converges weakly along a subsequence $\eps_k\to 0$
  to a distribution $W(t,\bx,\bk)\in {\mathcal C}^0(0,T;{\mathcal
  A}'(\Rm^{2d}))$.  } 

The proof of this lemma is obtained by rewriting $W_\eps$ as
\begin{displaymath}
  W_\eps(t,\bx,\bp)=\dint_{\Rm^{d}} \hat f(\bq)
  \dint_{\Rm^d}  e^{i\bp\cdot\bz}
   \tilde G_\eps(t,\bx-\dfrac{\eps\bz}{2};\bq) 
   \tilde G_{\eps*}(t,\bx+\dfrac{\eps\bz}{2};\bq) \dfrac{d\bz}{(2\pi)^d}d\bq,
\end{displaymath}
where $\tilde G_\eps(t,\bx;\bq)=\int_{\Rm^d}
G(t,\bx;\by)\chi_\Omega(\by)e^{-i\bq\cdot\bx/\eps}d\by$ solves
(\ref{eq:syst}) with initial data
$\chi_\Omega(\bx)e^{-i\bq\cdot\bx/\eps}I$, and $\tilde
G_{\eps*}(t,\bx;\bq)=\int_{\Rm^d}
G_*(t,\bx;\by)\chi_\Omega(\by)e^{i\bq\cdot\bx/\eps}d\by$ solves
(\ref{eq:greenadj}) with initial data
$\chi_\Omega(\bx)e^{i\bq\cdot\bx/\eps}A^{-1}(\bx)$.  The functions
$\tilde G_\eps$ and $\tilde G_{\eps*}$ are uniformly (in $\eps$ and
$\bq$) bounded in $L^2$, and hence \cite{GMMP,LP} $\int_{\Rm^d}
e^{i\bp\cdot\bz} \tilde G(t,\bx-\frac{\eps\bz}{2};\bq) \tilde
G_{*}(t,\bx+\frac{\eps\bz}{2};\bq) \frac{d\bz}{(2\pi)^d}$ is uniformly
bounded in ${\mathcal C}^0(0,T;{\mathcal A}'(\Rm^{2d}))$.  This implies the
result of the Lemma. \newline The Wigner distribution at time
$t=0$ is given by $W(0,\vx,\bk)=|\chi_\Omega(\vx)|^2\hat f(\bk)
A^{-1}(\vx)$.
The dispersion matrix $L(\vx,\bk)=A^{-1}(\vx)k_jD^j$ has a double eigenvalue
$\omega_0=0$ that corresponds to vortical modes, and simple eigenvalues
$\omega_{1,2}=\pm c(\vx)|\bk|$, $c(\vx)=1/\sqrt{\rho(\bx)\kappa(\bx)}$.
As was shown in \cite{GMMP,RPK-WM} the limit Wigner distribution
may be decomposed as 
\begin{equation}\label{wig-decomp}
W(t,\vx,\bk)=\sum_{j=1}^2 a_{ij}^0(t,\vx,\bk)\vb_i^0\vb_j^{0*}+
a_1(t,\vx,\bk)\vb^1\vb^{1*}+a_2(t,\vx,\bk)\vb^2\vb^{2*}.
\end{equation}
Here $\vb$ are eigenvectors of the matrix $L$. If the source $\bS(\vz)$
has no vortical waves, that is, if $\hat \bS(\bk)$ is parallel to $\bk$,
then the limit of the back-propagated signal is
\begin{eqnarray}\label{limit-back}
\bu^{B}(\Bxi; \bx_0)&=& \sum_{m=1}^2\dint_{\Rm^{2d}}
a_m(t,\vx_0,\bk)\Gamma \vb^m(\bx_0,\bk)\vb^m(\vx_0,\bk)
e^{-i\bk\cdot(\bz-\Bxi)}A(\bx_0)\Gamma\bS(\bz) d\bz d\bk\\
&=&
\dint_{\Rm^d} F(t,\Bxi-\bz;\vx_0)\bS(\bz)d\bz = (F(t,\cdot;\bx_0)*\bS)(\Bxi),
\label{eq:reconst}
\end{eqnarray}
with $a_1(0,\vx,\bk)=a_2(0,\vx,\bk)=|\chi_\Omega(\vx)|^2\hat f(\bk)$.
The quality of the refocusing of the back-propagated signal is determined 
by the decay properties in $\Bxi$ of the kernel 
\begin{equation}\label{eq:kernelF}
F(t,\Bxi;\vx_0)=\sum_{m=1}^2\dint_{\Rm^{d}}
a_m(t,\vx_0,\bk)\Gamma \vb^m(\bx_0,\bk)\vb^m(\vx_0,\bk)
e^{i\bk\cdot\Bxi}A(\bx_0)\Gamma d\bk.
\end{equation}
In the limit where $f=\delta$ and $\Omega=\Rm^d$, we verify that
$\bu^B(\Bxi;\bx_0)=\bS(\Bxi)$.  When all the information is propagated
back, the refocusing is perfect, as is physically expected.  However,
when the medium is homogeneous and $c(\vx)=c_0$, the amplitudes
$a_{1,2}(t,\vx_0,\bk)=|\chi_\Omega(\vx_0\mp c_0\hat\bk t)|^2\hat
f(\bk)$ become more and more singular in $\bk$ as time grows (their
gradient in $\bk$ grows linearly in time). The corresponding kernel
$F=F_H$ given by (\ref{eq:kernelF}) being a Fourier transform in $\bk$
of a function with growing gradients decays more and more slowly (in
$\Bxi$) as time grows. This leads to poorer and poorer refocusing.

%

\section*{\large 3. Refocusing via the Liouville and Fokker-Plank equations}

In this section, we analyze the random geometric acoustics limit,
where the correlation length $\delta$ of the random medium is small
compared to the propagation distance $L$ 
but large relative to the wavelength $\eps$. 
The mode $a=a^1$ satisfies the Liouville equation
\begin{equation}
  \label{eq:liouville}
  \partial_t a^\delta+\nabla_\bk\omega_\delta \cdot \nabla_\vx a^\delta-
\nabla_\vx\omega_\delta \cdot \nabla_\bk a^\delta=0,
\end{equation}
where $\omega(\vx,\bk)=(c_0+\sqrt{\delta}c_1(\vx/\delta))|\bk|$.  The
Liouville equation is obtained thanks to the a priori estimate of
Lemma 1.  We assume that $c_1(\bx)$ is a mean-zero stationary random
process with a smooth rapidly decaying correlation function $R(\vx)$.
In the limit $\delta\to 0$ we have 
${\mathbb E}\left\{a^\delta\right\}\to a(t,\bx,\bk)$, where $a$ solves the
Fokker-Plank equation \cite{Kesten-Papanico}
\begin{equation}
  \label{eq:Fokker-Plank}
  \pdr{a}{t} = 
\pdr{}{k_m}\left(k^2D_{jm}(\hat\bk)\pdr{a}{k_j}\right)-
c_0\hat\bk\cdot\nabla_\bx a 
\end{equation}
with the initial data $a(0,\vx,\bk)=|\chi_\Omega(\vx)|^2\hat f(\bk)$.
Here, $k=|\bk|$, and the diffusion matrix is
\begin{displaymath}
D_{jm}(\hat\bk)=
\frac{-1}{2}\int_{-\infty}^\infty\frac{\partial^2 R}{\partial z_j\partial z_m}
(c_0\hat\bk s)ds,
\end{displaymath}
with $\hat \bk=\bk/k$. The operator on the right side of
(\ref{eq:Fokker-Plank}) is degenerate because $D_{ij}(\bk)k_j=0$.
However, it is hypoelliptic on $X={\mathbb R}_\vx^3\times {\mathbb
  S}_\bk^2$ and thus possesses a smooth Green's function
$G_L(t,\bx,\bk;\bx',\bk')$ on $X$ so that
\begin{displaymath}
a(t,\vx,\bk)=\int_{\Rm^{2d}} G_L(t,\vx,\bk;\vy,\hat\bk')|\chi_\Omega(\vy)|^2
 \hat f(\hat\bk',k') d\vy d\bk'.
\end{displaymath}
Hence we can introduce the new filter $F_L=F_L^1+F_L^2$ with
\begin{equation}
  \label{eq:FLt}
 F_L^1(t,\Bxi;\bx_0)=\dint_{\Rm^{d}}e^{i\bk\cdot\Bxi}
a(t,\bx_0,\bk)\Gamma\vb^1(\bk)\vb^{1*}(\bk)A_0\Gamma  d\bk,
\end{equation}
and $F_L^2$ defined similarly. In the limit $\eps\to0$ and
$\delta\to0$ we obtain the reconstruction (\ref{eq:reconst}) for
${\mathbb E}\left\{u^B\right\}$ with $F=F_L$.  Now the function
$a(t,\vx,\bk)$ is smoother in $\bk$ than in the homogeneous case and thus the
filter $F_L$ decays faster than $F_H$.  This produces sharper
refocusing and eliminates spurious Fresnel zones.
The process $a^\delta(t,\vx,\bk)$ does not converge to the
deterministic function $a(t,\vx,\bk)$ in probability pointwise in
$\vx$ and $\bk$. However, for any test function $\phi(\vx,\bk)\in{\mathcal
  S}({\mathbb R}^{2d})$ the process $\phi^\delta(t)=(a^\delta,\phi)$
converges to $(a,\phi)$ in probability. That means that the refocused
signal $u^B$ is deterministic. This self-averaging of the refocused
pulse follows from slight modifications of the analysis in
\cite{Kesten-Papanico,PRS}.


\section*{\large 4. Refocusing via the transport and diffusion equations}

The radiative transport regime arises when the random medium has weak
fluctuations at the wavelength scale. This case is more complicated
than the one treated in Section 3 because the waves now fully interact
with the medium.  Propagation of wave correlations is described in
terms of a Boltzmann-type radiative transport equation
\cite{Erdos-Yau2,RPK-WM,Spohn}. The random fluctuations of the density
and compressibility are assumed to be of the form
\begin{math}
\rho_\eps(\bx)=\rho_0 + \sqrt\eps \rho_1(\frac{\bx}{\eps}),
\end{math}
\begin{math}
\kappa_\eps(\bx)=\kappa_0 + \sqrt\eps \kappa_1(\frac{\bx}{\eps}),
\end{math}
where $\rho_0$, $\kappa_0$ are constants, and $\rho_1(\bx)$,
$\kappa_1(\vx)$ are mean-zero stationary random processes. A rigorous
derivation of the transport equation for the Wigner distribution is
only available for Schr\"odinger equations \cite{Erdos-Yau2,Spohn},
Based on the formal analysis in \cite{RPK-WM} for hyperbolic systems,
we can still conjecture that $a_{1,2}(t,\vx,\bk)$ solve a linear
transport equation, at least after ensemble averaging over
realizations of the random fluctuations.  It is moreover known that
for long distances of propagation and large times, the transport
solution becomes independent of the direction $\hat\bk$ and is
approximated by that of a diffusion equation:
\begin{equation}
  \label{eq:diffusion}
  \partial_t a(t,\bx,k) - D(k)\Delta_\bx a(t,\bx,k)=0, 
\end{equation}
where $D(k)$ is a diffusion coefficient that depends on the power
spectrum of the random fluctuations (see \cite{RPK-WM} for instance).
When the filter $f$ is radially symmetrical, the solution of
(\ref{eq:diffusion}) may be written as
$a_{1,2}(t,\vx,k)=\psi(t,\vx,k)\hat f(k)$.  Here $\psi(t,\vx,k)$ is
the solution of (\ref{eq:diffusion}) with initial data
$\psi(0,\vx,k)=|\chi_\Omega(\vx)|^2$.  We then obtain that the new
filter $F_B$ is scalar and is given by
\begin{equation}
  \label{eq:Ft}
F_B(t,\Bxi;\bx_0)= \dint_{\Rm^d} e^{i\bk\cdot\Bxi}
\psi(t,\bx_0,k) \hat f(k) d\bk.
\end{equation}
The retransmitted field (in the weak sense and after ensemble
averaging) is given now by (\ref{eq:reconst}) with $F=F_B I$.  Note
that similarly to the random geometrical optics case treated in
Section 3 the amplitude $a(t,\vx,\bk)$ is smoother in $\bk$ than in
the homogeneous case and the filter $F_B$ decays faster than $F_H$,
thus produces a sharper refocusing than in a homogeneous medium.

{\bf Acknowledgement.} {We thank G. Papanicolaou and K.
Solna for fruitful discussions. G. Bal was supported by NSF grant
DMS-0072008 and L. Ryzhik by NSF grant DMS-9971742.}

%
\end{document}